\documentclass[12pt]{iopart}
\usepackage{graphicx} 
\usepackage[english]{babel}
\usepackage[utf8]{inputenc}
\usepackage{xcolor}

\usepackage{gensymb}
\begin{document}

\title[A home-lab experiment: resonance and sound speed] {A home-lab
  experiment: resonance and sound speed using telescopic vacuum
  cleaner pipes} 
\author{Martín Monteiro$^{1,2}$, Cecilia Stari$^2$, Arturo C. Mart{\'i}$^2$ }

\address{$^1$ Universidad ORT Uruguay, Montevideo, Uruguay}
\address{$^2$ Instituto de F\'{i}sica, Universidad de la
 Rep\'{u}blica, Montevideo, Uruguay}


\date{\today}
\begin{abstract}
We propose a home laboratory in which a telescopic vacuum cleaner pipe
and a smartphone are used to investigate sound speed and acoustic
resonance.  When the pipe is hit or the hands clapped near one end, the
sound produced is registered by a smartphone. The resonant frequency 
is obtained using a smartphone and an  appropriate application.
Varying the pipe's length and registering the corresponding resonant frequency
allows to obtain the sound speed. This home-lab, first proposed during
covid19 pandemic, has been incorporated as a home challenge to experiment
with acoustic resonance in new normal times.
\end{abstract}

\maketitle

\textbf{Resonances in telescopic pipes} When we listen to music coming
from a wind instrument, such as a flute or clarinet, we are witnessing
a phenomenon known as acoustic resonance. The instrumentalist injects
air into the edge of a hole, or through a mouthpiece, or a reed,
depending on the type of instrument, and the resulting vibrations
generate waves of various frequencies in the air inside the
instrument. According to the shape of the tube, some specific
frequencies, known as resonance frequencies, acquire predominant
energies, which determine the characteristic musical sound.  The
lowest frequency, also called the fundamental one, corresponds to the
musical note, while the highest frequencies, called
\textit{harmonics}, are responsible for the timbre or characteristic
sound of the instrument \cite{kinsler2000fundamentals}.

Let us consider a cylindrical pipe of length $L$ open at both ends. In
this case, the pressure at these points remain constant, so, they
behave as pressure nodes.  Taking into account this border condition,
the only stationary waves that can resonate in the tube are those in
which two of its nodes match both ends. Since the distance between
nodes is equal to half a wavelength, $\lambda$, then for resonant
waves it must be true that the length of the tube coincides with an
integer number of half-wavelengths: $L= 2 \lambda n$ where $n$ is a
natural number called a harmonic number. When perturbing abruptly the
pipe, for example hitting it with a blunt objetct, the fundamental
frequency which corresponds to $n=1$, prevails over the
harmonics. When varying the length of the pipe, the resonante
frequency, $f=c/\lambda$, where $c$ is the sound speed, is related to
the length $L$ by the following relationship $f=2 c / L$.

\textbf{The home experiment.}  In recent years, due to restrictions
related to covid19 pandemic, several laboratory activities were
modified to be proposed as home-labs \cite{o2021guide}.  A simple
experiment can be performed producing resonance inside a telescopic
vacuum cleaner pipe for different lengths. 
There are many ways to generate sound with a tube. 
Here we propose to do it with a small stroke of the hand. 
One end of the tube is placed near the smartphone and the opposite is tapped giving a sharp blow with the palm of the hand. At that moment you can see the frequency of the fundamental harmonic on the smartphone screen as a peak graph with the frequency value next to it (see right panel in Fig.\ref{fig2}). As this harmonic fades quickly, the pause function of the app must be used, to record the resonance frequency.

\begin{figure}[h]
\begin{center}
\includegraphics[width=0.6\textwidth]{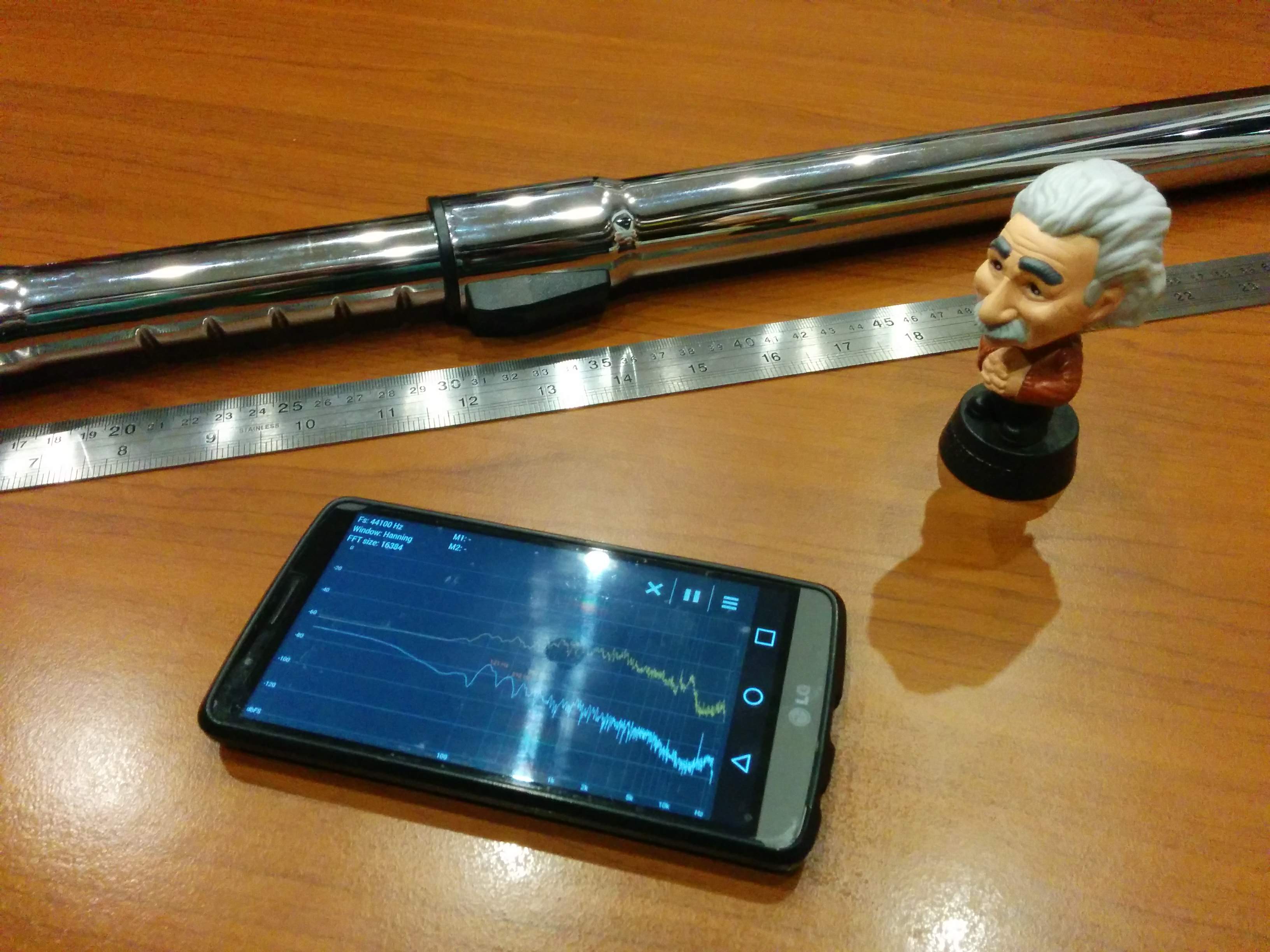}
\caption{The material employed in this experiment: an telescopic
  vacuum cleaner pipe, a ruler and a smartphone.}
\label{fig1}
\end{center}
\end{figure}

\begin{figure}[h]
\begin{center}
\includegraphics[width=0.265\textwidth]{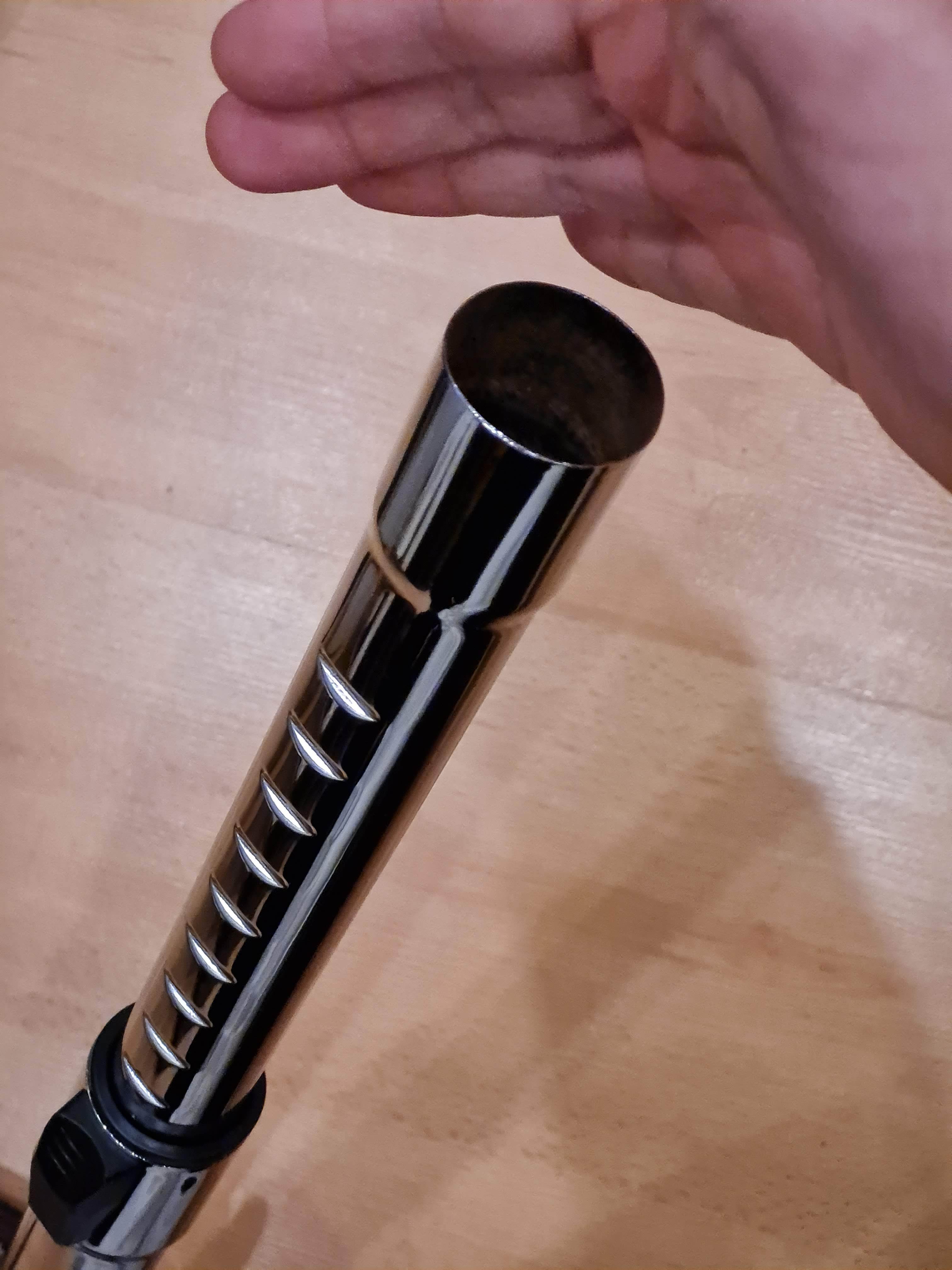}
\includegraphics[width=0.628\textwidth]{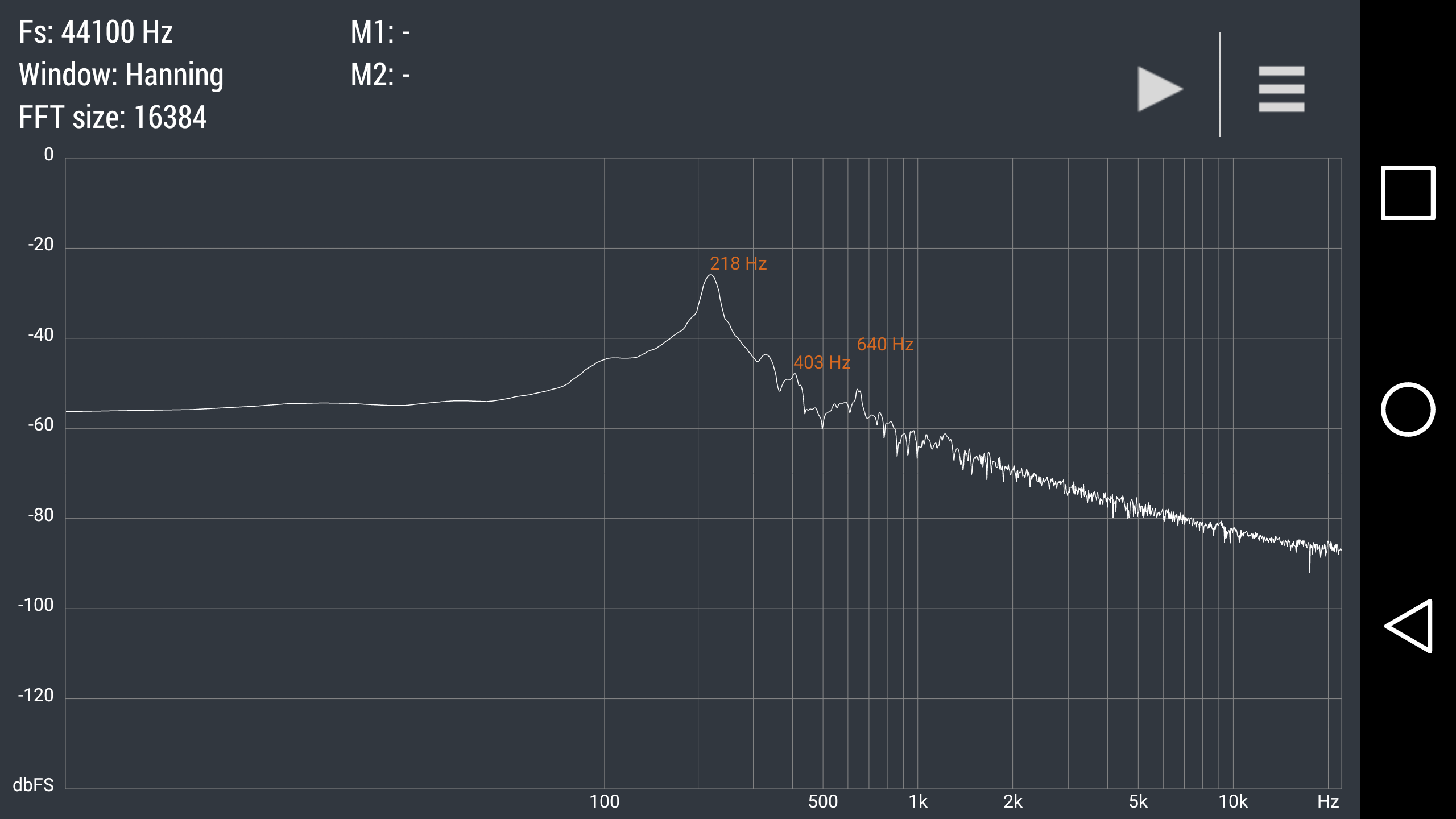}
\caption{The left panel shows the way to produce the sound in the tube while the right panel exhibits a screenshot of the Advanced Spectrum app whit the spectrum of one of the measurements. The horizontal axis is frequency (Hz)
and the vertical axis is energy (decibel). The main peak is the
fundamental frequency measured for the tube with length 79.8 cm.
}
\label{fig2}
\end{center}
\end{figure}

The resonance frequency can be measured very simply and economically
using smartphones sensors
\cite{vogt2012determining,yavuz2015measuring,monteiro2015measuring}
and one of the many free applications that analyze sound, such as
Physics Toolbox, Phyphox, Spectroid or Advanced Spectrum
\cite{monteiro2022resource}.  These apps (and others) use the great
computing power of these pocket computers to perform a Fast Fourier
Transform in real time and thus determine the frequencies present in
the sound that reaches the microphone of the phone. In general, to
perform a specific measure, it is necesseary to freeze the values
displayed on the screen.  In Fig.~\ref{fig1}, we show the experimental
setup and the few elements needed to perform this proposal.

\begin{figure}[th]
\begin{center}
\includegraphics[width=0.8\textwidth]{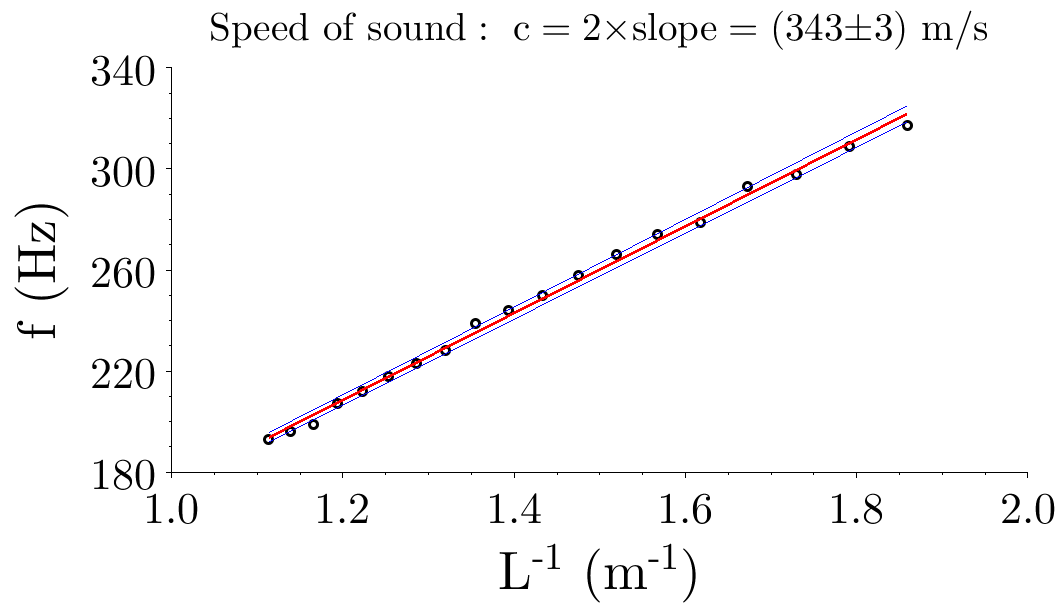}
\caption{Resonance frequency as a function of the inverse of the
  length.  The lineal fit indicated in the legend allows to obtain the
  sound speed.  }
\label{fig03}
\end{center}
\end{figure}

\textbf{Results and discussion.}  The screenshot of the Advanced
Spectrum app allows to obtain the spectrum of the sound as shown in
Fig.~\ref{fig2}. Repeating this measurement for different values of
the length of the telescopic tube it is possible to register the
resonant frequency for each length. In Fig.~\ref{fig03}, we plot the
frequency as a function of the inverse of the length of the tube and
perform a linear fit whose slope corresponds to half the speed of
sound. In this case, $c=(343 \pm 3)m/s.$ This value of the sound speed
agrees very well with the expected for the temperature at which the
experiment was performed, that is $22 \degree$ C. Theoretically the
speed of sound in a gas is, $c= \sqrt{\frac { \gamma RT } M }$, where
$\gamma$ for air is $1.4$, $R$ is the universal gas constant ($8.31$
J/K.mol) and $M$ is the molar mass, which for air is, $M$=0.029kg/mol
\cite{kinsler2000fundamentals}. With these values the expected speed
of sound at $22 \degree$ C is $ c=344.0$ m/s in very good agreement
with the experimental value.

\textbf{Final remarks} Recent covid19 pandemic obliged to rethink
the way we teach and learn physics.  Home activities have adquired
considerable  importance larger than before.  This proposal provides an
interesting opportunity to experiment acoustic resonance and sound
speed with everyday elements. Several extensions could be presented in
particular activities about Helmholtz resonators
\cite{monteiro2018bottle}, stationary waves in standard drain pipes
\cite{kasper2015stationary}, Kundts' tubes \cite{hellesund_2019} or
rthe measurements of pressure profiles \cite{Soares_2022} among many
others.

\textbf{Acknowledgments} We acknowledge financial support from grant
Fisica Nolineal (ID 722) Programa Grupos I + D CSIC 2018 (UdelaR,
Uruguay) and PEDECIBA (UdelaR, MEC, Uruguay).

\section*{References}

\bibliography{/home/arturo/Dropbox/bibtex/mybib}

\end{document}